\documentclass[fleqn,usenatbib]{mnras}

\usepackage{newtxtext,newtxmath}

\usepackage[T1]{fontenc}

\DeclareRobustCommand{\VAN}[3]{#2}
\let\VANthebibliography\thebibliography
\def\thebibliography{\DeclareRobustCommand{\VAN}[3]{##3}\VANthebibliography}

\usepackage{graphicx}	
\graphicspath{ {./Figures/} }
\usepackage{amsmath}	

\title[Blue Lurkers in NGC\,362]{GlobULeS-V. UVIT/{\it AstroSat} studies of  stellar populations in NGC\,362: Detection of Blue Lurkers in a Globular Cluster}

\author[Dattatrey et al.]{Arvind K. Dattatrey$^{1,2}$\thanks{arvind@aries.res.in}, 
R. K. S. Yadav$^1$\thanks{rkant@aries.res.in}, 
Gourav Kumawat$^{3}$,
Sharmila Rani$^{4}$,
Gaurav Singh$^{4}$,
\newauthor Annapurni Subramaniam$^{4}$\thanks{purni@iiap.res.in},
Ravi S. Singh$^{2}$
\\
$^{1}$Aryabhatta Research Institute of Observational Sciences, Manora Peak Nainital, 263001, India., arvind@aries.res.in\\
$^{2}$ Deen Dayal Upadhyay Gorakhpur University, Gorakhpur, Uttar Pradesh 273009, India.\\
$^{3}$Indian Institute of Science Education and Research, Bhopal 462066, India.\\ 
$^{4}$Indian Institute of Astrophysics, Koramangala, Bangalore 560034, India.\\
}
\date{Accepted 2023 May 12. Received 2023 May 6; in original form 2023 March 7}
\pubyear{2022}

\begin{document}
\label{firstpage}
\pagerange{\pageref{firstpage}--\pageref{lastpage}}
\maketitle

\begin{abstract}
We report the discovery of four blue lurkers with low- and extremely low-mass white dwarf (ELM WDs) companions in the Galactic globular cluster NGC\,362 using {\it AstroSat}'s Ultra Violet Imaging Telescope (UVIT). We analyzed the multi-wavelength spectral energy distribution (SED) of FUV-bright MS stars using data from the UVIT, UVOT, GAIA EDR3, and 2.2m ESO/MPI telescopes. Two each of low-mass WDs and ELM WDs are found as companions for the four blue lurkers by the fitting of two-component SED models. The effective temperatures, radii, luminosities, and masses of two low-mass WDs are (35000, 23000) K, (0.04, 0.05) R$_{\odot}$, (1.45, 0.22) L$_{\odot}$, and (0.2, 0.2)  M$_{\odot}$, while the two ELM WDs are (14750, 14750) K, (0.09, 0.10) R$_{\odot}$, (0.34, 0.40) L$_{\odot}$, and (0.18, 0.18)  M$_{\odot}$. The position of blue lurkers within the cluster shows that they originated via the Case A/B mass-transfer mechanism in a low-density environment. This is the first detection of blue lurkers with low-mass WDs and ELM WDs as companions in a globular cluster. The companion's cooling age is less than 4 Myr, which suggests that they were just recently formed. These binary systems might have originated due to the cluster's recent core collapse.
 
\end{abstract}

\begin{keywords}
ultraviolet: stars — (stars:) blue lurkers — (stars:)  Hertzsprung-Russell and CM diagrams — (stars:) white dwarfs — (Galaxy:) Globular star clusters: individual: (NGC362)
\end{keywords}

\section{Introduction} \label{sec:intro}
\cite{1953AJ.....58...61S} discovered stars brighter and bluer than the main sequence lying above the turn-off in the colour-magnitude diagram (CMD) of the Galactic Globular Cluster (GGC) M3. These stars are called Blue Straggler Stars (BSSs). The three major formation pathways for BSSs are as follows: mass transfer in binary star systems via Roche lobe overflow (RLOF), which leads to mass gain by BSS progenitors \citep{1964MNRAS.128..147M}; stellar collisions in a dense cluster environment \citep{1976ApL....17...87H}; and tightening and merger of inner binaries in a hierarchical triple system \citep{2009ApJ...697.1048P, 2014ApJ...793..137N}.\\ 
    Blue Lurkers (BLs), like BSSs, are also post-mass transfer (MT) systems, but they do not appear brighter than the main-sequence turn-off (MSTO) in the CMD. This could be due to a lack of accreted matter or a too-small accretor star. The jump in the CMD for these stars is insufficient for them to be seen above the MSTO. As a result, these stars have been designated as blue lurkers: “BSS-type stars lying hidden in the MS below MSTO” \citep{2019ApJ...881...47L}. These stars cannot be detected using the CMD alone because they do not stand out like BSSs. Some possible techniques to find these stars are as follows: Observation of more than average rotation (Vsini) indicating a recent MT event \citep{2019ApJ...881...47L}; Multi-wavelength SED analysis to find companions such as Extremely Low Mass (ELM) White Dwarfs (WDs), hot sub-dwarfs etc. \citep{2019ApJ...886...13J, 2020JApA...41...45S}; Looking for chemical peculiarities possible only via mass accretion. In fact, N-body simulations and population synthesis studies predict that such mass transfer products could be abundant \citep{2006ApJ...646.1160A,2013AJ....145....8G}.
 
ELM WDs are helium core WDs with masses M$\le$ 0.18 M$_{\odot}$ \citep{2018ApJ...858...14S}. The universe's age limits the mass of WDs formed by isolated star evolution to $\approx$ 0.4 M$_{\odot}$\citep{2010ApJ...723.1072B}. WDs with masses 0.1-0.4 M$_{\odot}$ were found to be part of compact binaries \citep{1995MNRAS.275..828M, 2004MNRAS.352..249B, 2010ApJ...723.1072B}. Their formation can be explained by mass loss in binary systems.

Relatively few studies have been done to identify BLs, and only in open clusters (OCs) \citep{2019ApJ...881...47L, 2019ApJ...886...13J, 2020JApA...41...45S}. These BLs are found in binary systems with WD/low-mass WD  companions. To our knowledge, no literature exists on the detection of BLs in GCs. Here we report the first detection of 
BLs in the GC NGC\,362 using UV data.

\begin{figure*}
\hspace*{-0.65cm}
\label{fig:CMD.png}
\includegraphics[width=5.0cm, height=5.0cm]{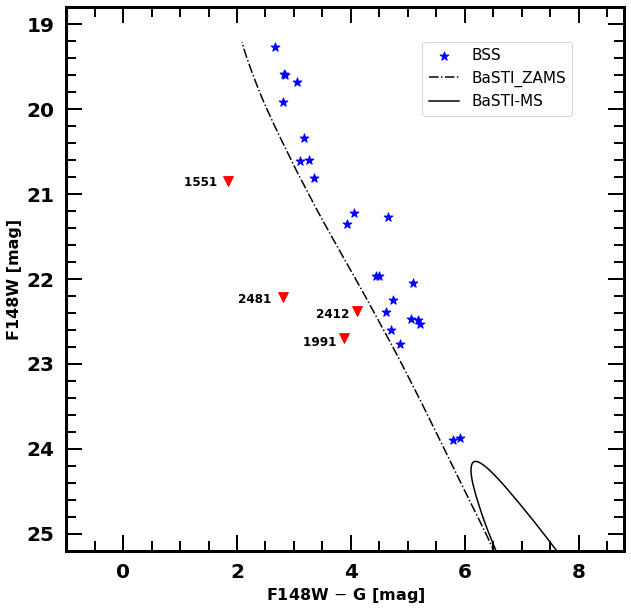} \label{fig:CMD_gaia}
\includegraphics[width=5.0cm, height=5.0cm]{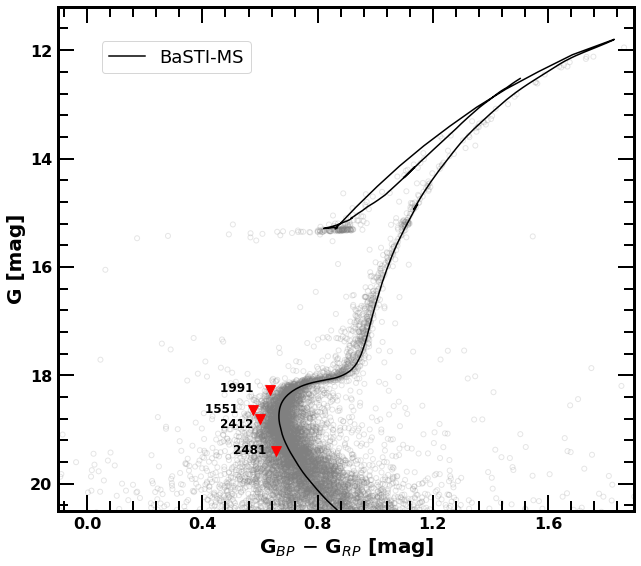}
 \includegraphics[width=5.0cm, height=5.0cm]{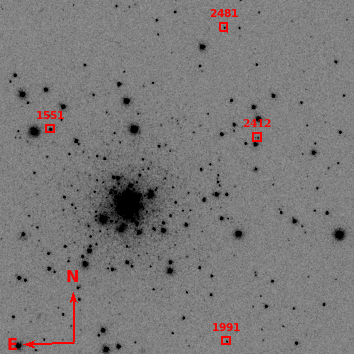}

\caption{Left panel: UV-optical CMD (F148W vs F148W $-$ G) of the BSSs for the cluster NGC\,362. Blue symbols represent the BSSs. The probable four BL candidates, bluer than the ZAMS, are displayed with red down-triangle symbols.\label{fig:FUV_G_CMD}  Middle panel: The optical CMD (G vs G$_{BP}$ $-$ G$_{RP}$) of the cluster NGC\,362. The BaSTI isochrone is overplotted with a solid black line. The four red down-triangle symbols represent the FUV bright MS stars. Right panel: Spatial locations of the four FUV bright MS stars in F148W image of NGC\,362. The field of view of the image is $\sim$8.5$'$ x 8.5$'$. North is up and east is on the left.}

\end{figure*}

 
The GC NGC\,362 is located in the Tucana constellation in the Southern hemisphere
(RA = $01^h\, 03^m\, 14.26^s$ and Dec = $-70^{\circ}$  50$\arcmin$ 55$\farcs$6). The cluster has an age of $\approx$ 11 Gyr, located at a distance of $\approx$ 8.83 kpc \citep{2021MNRAS.505.5978V}, a metallicity \big[Fe/H\big] of $\approx$ $-$1.3 dex, and a reddening of $\sim$ 0.05\,mag \citep{2010arXiv1012.3224H}.

\section{DATA SETS AND THEIR REDUCTION}
\label{sec:data}
The observations of NGC\,362 were made with the UVIT on November 11, 2016, in four UV filters: F148W, F169M, N245M, and N263M. UVIT is one of the payloads on {\it AstroSat}, India's first multi-wavelength space observatory, which launched on September 28, 2015. The exposure times were 4900, 4600, 4900, and 4600 sec in the F148W, F169M, N245M, and N263M filters, respectively. Using CCDLAB software \citep{2017PASP..129k5002P}, which corrects for satellite drift, flat-field, geometric distortion, fixed pattern noise, and cosmic rays, data reduction of the raw images was carried out. Detailed descriptions of the telescope, instruments and preliminary calibration can be found in \cite{2016SPIE.9905E..1FS} and \cite{2017AJ....154..128T}.

The archival UV data collected with Ultraviolet Optical Telescope (UVOT) are also used in this analysis. The raw data from the HEASARC archive\footnote{https://heasarc.gsfc.nasa.gov} was processed using the HEA-Soft\footnote{https://heasarc.gsfc.nasa.gov/docs/software/heasoft} pipeline. The telescope's details and photometric calibration of UVOT data can be found in \cite{2008MNRAS.383..627P}. Corrections were made to exposure maps and auxiliary spacecraft data, as well as to science images that were geometrically corrected for sky coordinates. In order to process the UVOT/Swift data, the procedure outlined by \cite{2014AJ....148..131S} was used. 
In addition, archival optical data in the $U, B, V, R,$ and $I$ filters collected by the Wide-Field Imager (WFI) mounted on the 2.2m ESO/MPI Telescope were utilised. The details of data reduction and photometry are provided in \cite{2006A&A...454.1029A}.

\begin{figure*}
\includegraphics[scale=0.415]{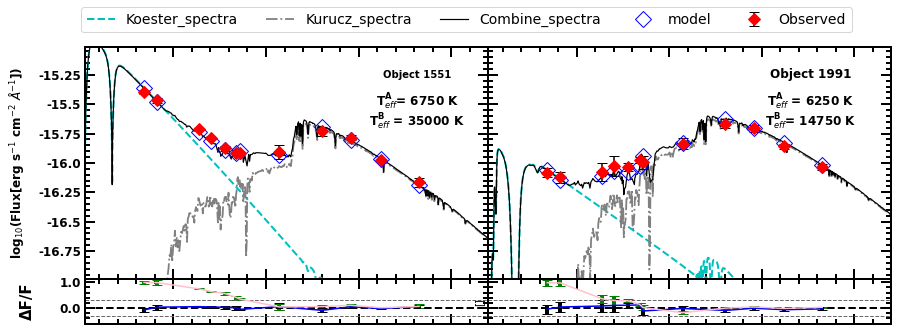}
\includegraphics[scale=0.425]{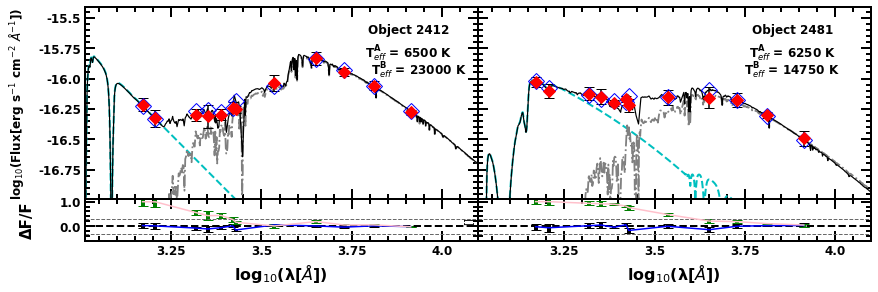}

\caption{The SEDs of four FUV bright MS stars (1551, 1991, 2412, and 2481). The T$_{eff}$ of the cool (A) and hot (B) components are displayed in each SED. The observed data points are represented by red solid-diamond points, while open blue diamond points represent the model points. Cyan, grey, and black curves represent the Koester, Kurucz, and combined spectra, respectively. $\Delta F/F$ is the fractional residual.}

\label{fig:SED}
\end{figure*}

\section{UV AND OPTICAL COLOUR MAGNITUDE DIAGRAMS}
\label{sec:cmds}
\subsection{The selection of FUV-bright stars on the main sequence}
Before identifying the FUV bright MS stars, we cross-matched the UVIT data with the \textit{{\it Gaia}} EDR3 photometric data with a matching radius of $1\farcs0$. The FUV bright MS stars are selected using the UV-optical CMDs. We utilised the proper-motion membership probability catalogue from the GlobULeS I paper \citep{2022MNRAS.514.1122S} to select the members.  The membership probabilities are based on the GAIA DR3 proper motions and a comprehensive comparison of the cluster like stars kinematics distribution and the field distribution. Stars with a membership probability greater than 85\% were chosen as cluster members and considered for further analysis. We constructed the UV-optical CMDs of members by finding the optical counterparts of the UVIT-detected sources.

We plotted the FUV optical CMD (F148W, (F148W $-$ G)) of BSSs shown in the left panel of Figure~\ref{fig:FUV_G_CMD}. The dotted and solid lines represent the ZAMS and MS curves taken from BaSTI\footnote{http://basti-iac.oa-abruzzo.inaf.it/}. The BSS (blue symbols) sequence is easily identified in the CMD. Four FUV bright stars (ID: 1551, 1991, 2412, and 2481) can be found bluer than the ZAMS. To check the location of these four FUV bright stars in the optical CMD, we constructed (G, (G$_{BP}$ $-$ G$_{RP}$)) diagram shown in the middle panel of Figure \ref{fig:CMD_gaia}. For visual guidance, we overplotted BaSTI isochrone with a solid line using the cluster's parameters given in Sec. \ref{sec:intro}. It is evident that the four FUV bright stars are located close to the turn-off point of the cluster in the optical CMD. We analysed the locations and contamination of these stars in the observed image and marked them with red squares in the F148W image, as depicted in the right panel of Figure \ref{fig:CMD_gaia}. This figure reveals that the stars are clearly resolved, with no contamination. They are situated more than 3$\arcmin$ from the cluster's centre.

\section{Spectral Energy Distributions}

In this section, we build the spectral energy distribution (SED) of the FUV bright MS stars 1551, 1991, 2412 and 2481 to look for the companions and their properties. We used the virtual observatory tool, VOSA (VO SED Analyzer, \citep{2008A&A...492..277B}) for this. VOSA uses filter transmission curves to compute synthetic photometry for a chosen theoretical model. The synthetic photometric data is then compared to the observed data, and the best fit is determined using the $\chi^2$ minimization test.  The specific procedures for single and binary component SED fitting are provided in \cite{2016ApJ...833L..27S} and \cite{2021MNRAS.501.2140R}.

In brief, we fit the model of \cite{2003IAUS..210P.A20C}  of varying temperature and surface gravity with fixed metallicity of \big[Fe/H\big]= -1 dex using the basic parameters provided in Sec. \ref{sec:intro}. The effective temperatures were considered to be in the 5000 $-$ 50000 K range, with log\,{\it g} ranging from 3$-$5 dex.
UVIT data were combined with UVOT and ESO 2.2m optical data for long coverage of wavelengths in the SED.

We used the Kurucz stellar atmospheric model to fit the single component SED, as shown with a grey curve in Figure \ref{fig:SED}. The observed data are represented by red points with error bars, whereas the model data is represented by blue points. The residual between the fitted model and the observed fluxes, normalized by the observed flux, are shown in the bottom panel of each SED. The green points, along with the error bars, are the residuals, and the dashed horizontal lines at ± 0.3 (30\%) are the residual threshold. The residual plots in Figure \ref{fig:SED} show that the residual flux in the UV region is greater than 0.3 in more than one data point. This indicates a UV excess in all four FUV bright MS stars, and a combination of hotter and cooler spectra may fit the SEDs. 

\begin{table*}
\centering
 \caption{The best-fit parameters of the cool and hot components. Here, T$_{eff}$ is the effective temperature in K, $\chi^2_r$ is reduced $\chi^2$, luminosity, radius, and mass are in the solar unit, V$_{gf}$ \& V$_{gfb}$ are the visual goodness of fit parameters, N$_{fit}$ is the total number of points taken in to account during the fitting. The mass and age of cool and hot companions are listed in the last two columns.}
\begin{tabular}{p{0.5cm} p{1.1cm} p{1.3cm} p{1.5cm} p{0.8cm} p{0.8cm} p{1.5cm} p{1.cm} p{0.7cm} p{0.5cm} p{0.85cm} p{0.7cm} p{0.82cm} } 
 \hline
    Name & RA       & DEC   & T$_{eff}$ (K)           &	Log g &	$\chi^2_r$ &	L/L$_{\odot}$ & R/R$_{\odot}$ &	V$_{gf}$ &	V$_{gfb}$ & N$_{fit}$ &Mass (M$_{\odot}$) &Age (Myr)\\ 
    \hline
1551A  &15.89396 &-70.81936 & 6750$^{+125}_{-125}$     & 3.0  & 58.76    & 2.45$^{+0.009}_{-0.300}$     & 1.15$^{+0.005}_{-0.002}$   & 29.8 & 1.37 & 12/12 &0.92&5500  \\ 
1551B  & &                  & 35000$^{+3000}_{-1000}$  & 9.5  &          & 1.45$^{+0.292}_{-0.170}$     & 0.04$^{+0.003}_{-0.005}$ &        &      &    &0.2&  $<0.10$   \\   
\\
1991A  &15.69960 &-70.90077 &6250$^{+125}_{-125}$     & 3.5   & 6.07    & 3.21$^{+0.181}_{-0.020}$    & 1.53$^{+0.001}_{-0.003}$   & 4.68 & 0.21 & 12/12  &0.85&10000  \\
1991B  & & &                14750$^{+250}_{-500}$     & 7.0   &          & 0.34$^{+0.0134}_{-0.010}$    & 0.09$^{+0.008}_{-0.003}$   &      &      &   &0.18 & $\sim4.0$ \\ 
\\
2412A  &15.65796 &-70.82486 & 6500$^{+125}_{-125}$     & 3.5  & 4.39     & 1.99$^{+0.294}_{-0.121}$     & 1.18$^{+0.001}_{-0.002}$   & 4.37 & 0.88 & 12/12    &0.87&8000   \\ 
2412B  & &                  &23000$^{+500}_{-1000}$    & 6.5  &          & 0.22$^{+0.011}_{-0.003}$  & 0.05$^{+0.005}_{-0.002}$ &      &      &     & 0.2& $<0.10$  \\ 
\\
2481A  &15.69260 &-70.78316 & 6250$^{+125}_{-125}$     & 5.0  & 15.25    & 1.02$^{+0.270}_{-0.080}$      & 0.90$^{+0.015}_{-0.015}$   & 10.70 & 2.92 & 12/12    &0.82&5500   \\
2481B  & &                  &14750$^{+250}_{-750}$     & 9.5  &          & 0.40$^{+0.085}_{-0.001}$   & 0.10$^{+0.013}_{-0.001}$    &      &       &        & 0.18 & $\sim4.0$   \\ 
\\

\hline
\end{tabular}
 \label{tab:parameters}
\end{table*}

We fitted two$-$component SEDs to account for the UV excess.
We used the Kurucz stellar atmospheric model \citep{1997A&A...318..841C,2003IAUS..210P.A20C} to fit the cooler component and the Koester WD model \citep{2010MmSAI..81..921K} (shown with cyan curves) to fit the hotter component. 
The T$_{eff}$  and log\,{\it g} in the Koester WD model range from 5000$-$80000 K and 6.5$-$9.5, respectively. The combined spectra are shown with a black line in Figure \ref{fig:SED}. The fitting parameters $\chi^2_r$, V$_{gf}$, and V$_{gfb}$ of the combined spectra, listed in Table \ref{tab:parameters}, indicate that the spectrum fits well for all wavelengths. The Koester model fits the hot components well, while the Kurucz model fits the cool components well. The fractional residuals for the combined SEDs are shown as black points along with the error in the bottom panels. For the combined SEDs, the residuals are within 0.3 on each wavelength.

Fitting the two components yields the fundamental parameters of the cool and hot components of the FUV bright MS stars. The estimated parameters are shown in Table \ref{tab:parameters}. The star IDs with "A" and "B" represent the cool and hot components, respectively. The range of parameters for cool components are T$_{eff}$ $\sim$ 6250 $-$ 6750 K, log\,{\it g} $\sim$ 3 $-$ 5, L $\sim$ 1.02 $-$ 3.21 L$_{\odot}$, and R $\sim$ 0.90 $-$ 1.53 R$_{\odot}$, while those for hot components are, T$_{eff}$ $\sim$ 14750 $-$ 35000 K, log $g$ $\sim$ 6.5 $-$ 9.5, L $\sim$ 0.22 $-$ 1.45 L$_{\odot}$, and R $\sim$ 0.04 $-$ 0.10 R$_{\odot}$. 

 The effective temperature and radius of 11 BLs found in the open cluster M67 have been calculated by \citet{2019ApJ...881...47L} using SED analysis. The effective temperature and radius were found to be in the 5520-6840 K and 0.85-1.9 R$_{\odot}$ ranges, respectively. The present estimated parameters for the BLs are well within the range.

\subsection{Age and mass of the cool and hot components}
\begin{figure}
\centering
\includegraphics[scale=0.2]{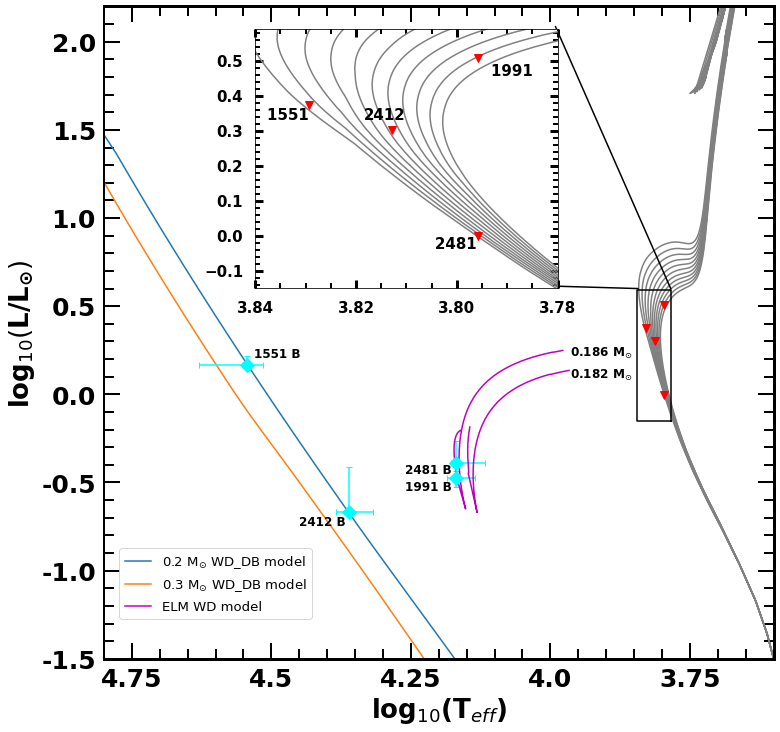}
\caption{The HR diagram of the cool and hot components. Red and cyan points, respectively, represent the cool and hot components. The grey, sky blue, and magenta curves represent the BaSTI isochrones, the white dwarf model, and the ELM models, respectively. The WD model is for 0.2 M$_{\odot}$, while the ELM model is for 0.182 and 0.186 M$_{\odot}$.}
\label{fig:HR}
\end{figure}

We estimated the age and mass of the cool components, 1551A, 1991A, 2412A, and 2481A, by comparing their positions to theoretical isochrones retrieved from  BaSTI (\cite{2021ApJ...908..102P}) in the H-R diagram shown in Fig. \ref{fig:HR}. We considered the SED-estimated surface temperature and luminosity of cool components, which are represented by red points in the figure. The isochrones from 5 to 11 Gyr are plotted with intervals of 0.5 Gyr using grey lines with the cluster's input parameters provided in Sec~\ref{sec:intro}. By comparing the positions to the isochrones, the cool components 1551A, 1991A, 2412A, and 2481A have ages of 5.5, 10, 8, and 5.5 Gyr, respectively, and masses of 0.92, 0.85, 0.87, and 0.82 M$_{\odot}$. The estimated age and mass of each cool component are displayed in Table \ref{tab:parameters}.  

The age and mass of 1551B, 1991B, 2412B, and 2481B (hot components) are derived using the H-R diagram depicted in Fig. \ref{fig:HR}. The surface temperature and luminosity derived using SEDs are used in the H-R diagram. Cyan points denote the hot components. A blue and yellow line represents a 0.2 and 0.3 M$_{\odot}$ helium DB WD models from \cite{2009ApJ...696.1755T}. The ELM WD models of mass 0.182 and 0.186 M$_{\odot}$ are shown with magenta lines, which were taken from \cite{2013A&A...557A..19A}. The two hot components, 1991B and 2481B are located on the 0.186 M$_{\odot}$ ELM WD model, while 1951B and 2412B are very close to the 0.2 M$_{\odot}$ WD DB model. The models predict cooling ages of $\leq$ 0.1 Myr for 1551B and 2412B and 4 Myr for 1991B and 2481B. The mass and cooling age of the hot components are listed in Table \ref{tab:parameters}.   

\section{Results and discussion}\label{sec:results}

Using the UVIT, UVOT, and 2.2m ESO optical data, we discovered four FUV bright MS stars in the GGC NGC\,362. We built the SEDs using VOSA to determine their physical characteristics and identify any companions that might be present. The fitting of single component SED to the observed data points reveals the UV excess in all four stars. To account for the UV excess, we fitted the SED of two components. The combined spectra of cool and hot components fit well to all four stars. This study indicates that these stars are in binary systems.

Based on the current analysis of the SEDs of four FUV bright MS stars, the surface temperature, luminosity, radius, mass and age of the cool components 1551A, 1991A, 2412A, and 2481A are determined to be in the range of 6250 - 6750 K, 1.02 - 3.21 L$_{\odot}$, 0.90 - 1.53 R$_{\odot}$, 0.82 - 0.92 M$_{\odot}$, and 5.5 - 10 Gyr. The masses of these components are near the cluster's turn-off mass (0.8 M$_{\odot}$). In comparison to the BSSs, their positions on the H-R diagram are close to the cluster's turn-off point. The derived age of the cool components reveals that they are younger than the cluster (11 Gyr). The present analysis leads us to believe that these cool components could be BLs. They have been regenerated by mass transfer like the BSSs but mixed with the normal MS population in the CMD. As the cluster ages, similar mass stars will evolve away from the MS, showing these lurkers as BSSs \citep{2019ApJ...881...47L}.

The surface temperatures, luminosities, and radii of the hot components 1551B, 1991B, 2412B, and 2481B were found to range from 14750 - 35000 K, 0.22 - 1.45 L$_{\odot}$, and 0.04 - 0.1 R$_{\odot}$, respectively, based on the best SED fits.

The mass and cooling age of 1551B and 2412B are 0.20 M$_{\odot}$ and $\leq$0.1 Myr, respectively, while those of 2481B and 1991B are 0.186 M$_{\odot}$ and 4 Myr. Based on the estimated parameters, locations, and superimposed models in the H-R diagram, we can deduce that 1551B and 2412B are low-mass Helium core WDs, while 2481B and 1991B are ELM WDs. 

We found that the newly discovered FUV bright MS stars are binary systems. The stars 1551 and 2412 have a pair of BL+low-mass WDs, whereas 2481 and 1991 have a pair of BL+ELM WDs. Because BLs are a smaller mass group of BSSs, the formation mechanism may be similar to that of the BSSs. Similar to BSSs, these are post-mass transfer systems. The masses of hot companions are found to be 0.186 and 0.2 M$\odot$. It is unlikely that such low-mass WD candidates will form through a single, isolated star evolution. They might have originated in binary systems. The mass of WDs created by a single star evolution is restricted to 0.4 M$\odot$.\citep{2010ApJ...723.1072B}. 
The age of the universe limits the lower end of the WD mass range. However, ELM WDs can be found in binaries \citep{2019ApJ...883...51R,2019ApJ...886...13J,2020JApA...41...45S,2022MNRAS.511.2274V,2023ApJ...943..130D}. Mass loss in the early phases of evolution sets a minimum value for WD masses. It is possible that the case A/B MT will lead to the He-core WDs through early envelope mass loss \citep{1991ApJS...76...55I,1995MNRAS.275..828M}. As a result, MT is required for the creation of low-mass WDs in tight binary systems where the companion tears away the low-mass WDs progenitor's envelope and the low-mass core fails to ignite the He core. The companion star acquired mass and evolved into BLs during the mass loss. \cite{2019ApJ...871..148L} claim that ELM WDs with masses less than 0.22 M$_{\odot}$ are generated through the Roche lobe overflow channel. This leads us to the further conclusion that these BLs are produced via MT with a WD companion.

The formation of investigated binary systems may be related to the dynamical evolution of the cluster. The cooling ages of low-mass WDs and ELM WDs are $<$ 0.1 and 4 Myr, respectively. \cite{2013ApJ...778..135D}  suggest that NGC\,362 is either experiencing or is about to experience the core collapse based on the density profile. The cooling ages of the hot companions show that they have only recently formed. As a result, we might infer that these binary systems might have originated during the cluster's core collapse. During the collapse, the centre density grows quickly, increasing the chances of gravitational interactions as well \citep{1997A&ARv...8....1M}; the creation of binary systems that are brought into the MT regime by hardening processes caused by gravitational encounters \citep{1990ApJ...362..522M,2008AJ....135.2129H}.  On the other hand, based on the radial distribution of BSS,  \cite{2013ApJ...778..135D} found that the core-collapse may have begun around 200 Myr ago, which is significantly earlier than the expected formation epoch of BLs. Therefore, presuming that cluster properties can play a role in the formation of BLs, it is more probable that they formed during the post-core-collapse rebounce phases.

The four BLs are located more than 3$^\prime$ from the cluster centre. The origin of these BLs in the cluster's outskirts is unknown. We speculate that these BLs, like BSS,  may have been created by collisions during resonance encounters between binary and single stars in the core and ejected into the outer regions by the recoil from the interactions \citep{1993ApJ...408L..89B, 1994ApJ...431L.115S}. The discovery of BLs + WDs systems in the outer area of the near-core collapse cluster has consequences for GCs' dynamical evolution theories. The total census of BLs on cluster MSs is mainly unknown because of the difficulty of discovering them. The stellar characteristics of this vital population are, as a result, poorly understood.

Another possibility for the origin of these BLs is the mass transfer in the primordial close binary systems located in the outer region of the cluster. According to \cite{2013ApJ...778..135D}, NGC\,362 is a dynamically old cluster, which means that even the external BSSs have sunk towards the cluster's centre due to the mass segregation process. As BLs have low mass compared to BSSs, the impact of mass segregation on BLs will be smaller than that on BSSs. As a result, they are still in the cluster's outer region. 

 The BLs are located in the outer part of the cluster, which makes them an intriguing spectroscopic study target. The spectroscopy can provide an accurate estimation of the parameters and their nature to constrain their formation paths.

 Due to the low spatial resolution, we do not explore the BLs at the cluster's centre. To get a complete sample of BLs, we need high resolution and deeper UV observations.

\section{Summary and Conclusions}\label{sec:concludions}

The first detection of BLs in the GGC NGC\,362 is presented. This result is based on observations from {\it AstroSat} UVIT as well as archival data from UVOT and the 2.2m ESO/MPI telescopes. We can draw the following conclusions from this study.

\begin{enumerate}

\item We found four FUV-bright MS stars in cluster NGC\,362. The SED study reveals an excess in FUV flux from all four sources. To account for the FUV excess, we fitted a two-component SED model.

\item The cool companions are fitted with the Kurucz model, and the hot companions are fitted with the Koester model. The parameters obtained from the two-component SED model are shown in Table \ref{tab:parameters}. Based on our SED analysis, we conclude that the cool companions of all four sources are BLs. The hot companions of 1551 and 2412 are low-mass Helium core WDs, whereas 2481 and 1991 are ELM WDs.

\item The location of these four sources within the cluster suggests they were formed via the Case A/B mass transfer process. The cluster is currently undergoing a core collapse. The cooling age of hot companions indicates that they were generated very recently. We suggest that these binary systems were formed during the core collapse process and were ejected from the core due to gravitational interaction. 

\end{enumerate}

\section*{Acknowledgements}
We warmly thank Pierre Bergeron for providing us with the WD cooling models generated for UVIT filters. I would like to thank Dr. Sindhu Pandey for her help with UVIT data reduction. I want to thank Gurpreet Singh for his assistance during this project. AS thanks the support of the SERB power fellowsip. This paper is based on observations collected at the European Organization for Astronomical Research in the Southern hemisphere under ESO program 60.A-9121(A).

\section*{DATA AVAILABILITY}
The data utilized in this article will be provided upon request.

\bibliographystyle{mnras}
\bibliography{references} 

\bsp	
\label{lastpage}
\end{document}